\documentclass[journal]{IEEEtran}
\ifCLASSINFOpdf
\else
\fi

\usepackage{cite}
\usepackage{amsmath,amssymb,amsfonts}
\usepackage{algorithmic}
\usepackage{graphicx}
\usepackage{mdframed}
\usepackage{textcomp}
\usepackage{subfigure}
\usepackage{caption}
\usepackage[table]{xcolor}
\usepackage{listings}
\usepackage{adjustbox}
\usepackage{arydshln}
\usepackage{flushend}

\usepackage{multirow}

\def\BibTeX{{\rm B\kern-.05em{\sc i\kern-.025em b}\kern-.08em
    T\kern-.1667em\lower.7ex\hbox{E}\kern-.125emX}}

\usepackage{color,soul}

\begin{document}
%
% paper title
% Titles are generally capitalized except for words such as a, an, and, as,
% at, but, by, for, in, nor, of, on, or, the, to and up, which are usually
% not capitalized unless they are the first or last word of the title.
% Linebreaks \\ can be used within to get better formatting as desired.
% Do not put math or special symbols in the title.
\title{DB4HLS: A Database of High-Level Synthesis \\ Design Space Explorations}
%
%
% author names and IEEE memberships
% note positions of commas and nonbreaking spaces ( ~ ) LaTeX will not break
% a structure at a ~ so this keeps an author's name from being broken across
% two lines.
% use \thanks{} to gain access to the first footnote area
% a separate \thanks must be used for each paragraph as LaTeX2e's \thanks
% was not built to handle multiple paragraphs
%

\author{\IEEEauthorblockN{Lorenzo Ferretti\textsuperscript{1}, \IEEEauthorblockN{Jihye Kwon}\textsuperscript{2}, Giovanni Ansaloni\textsuperscript{3}, Giuseppe Di Guglielmo\textsuperscript{2}, Luca Carloni\textsuperscript{2}, Laura Pozzi\textsuperscript{1} }\\
\IEEEauthorblockA{\textit{\textsuperscript{1}Universit\`a della Svizzera italiana, Lugano, Switzerland,} }
\IEEEauthorblockA{\textit{\textsuperscript{2}Columbia University, New York, United States}}\\
\IEEEauthorblockA{\textit{\textsuperscript{3}EPFL, Lousanne, Switzerland}}}
\maketitle

% For peer review papers, you can put extra information on the cover
% page as needed:
% \ifCLASSOPTIONpeerreview
% \begin{center} \bfseries EDICS Category: 3-BBND \end{center}
% \fi
%
% For peerreview papers, this IEEEtran command inserts a page break and
% creates the second title. It will be ignored for other modes.
%\IEEEpeerreviewmaketitle

% As a general rule, do not put math, special symbols or citations
% in the abstract or keywords.
\begin{abstract}

High-Level Synthesis (HLS) frameworks allow to easily specify a large number of variants of the same hardware design by only acting on optimization directives. Nonetheless, the hardware synthesis of implementations for all possible combinations of directive values is impractical even for simple designs. Addressing this shortcoming, many HLS Design Space Exploration (DSE) strategies have been proposed to devise directive settings leading to high-quality implementations while limiting the number of synthesis runs. All these works require considerable efforts to validate the proposed strategies and/or to build the knowledge base employed to tune abstract models, as both tasks mandate the syntheses of large collections of implementations. Currently, such data gathering is performed ad-hoc, a) leading to a lack of standardization, hampering comparisons between DSE alternatives, and b) posing a very high burden to researchers willing to develop novel DSE strategies. Against this backdrop, we here introduce DB4HLS, a database of exhaustive HLS explorations comprising more than 100000 design points collected over 4 years of synthesis time. The open structure of DB4HLS allows the incremental integration of new DSEs, which can be easily defined with a dedicated domain-specific language.  
We think that of our database, available at \small{\texttt{https://www.db4hls.inf.usi.ch/}}, will be
%extremely compelling 
%LP: maybe a bit too strong....
a valuable tool 
for the research community investigating automated strategies for the optimization of HLS-based hardware designs.

%High-Level Synthesis (HLS) tools allow the generation of a large variety of hardware implementations from the same specification by setting different optimization directives.
%While one hand  HLS design flows allow designers to easily explore different design alternative, they also require  strategies able to navigate the vast design space of microarchitectural alternatives to identify the most promising implementations while minimising the number of time-consuming HLS runs.
%It allows fast exploration of different design alternatives. 
%In the past years many research efforts have lead to different strategies aiming at identifying the most promising implementations while minimising the number of time-consuming HLS runs.
%Despite the many publication, the evaluation of such methodologies lacks in standardisation. Moreover, recently published works are showcasing the importance, for such methodologies, to rely on large datasets to build accurate models. Due to the long time required to generate such data, there is scarcity of such resources.
%We preset DB4HLS, a collection of HLS implementations gathered targeting the Machsuite benchmarks and a framework to expand the database. Our contributions aim at offering researchers the possibility to standardize the evaluation process of existing methodologies and an important source of data for ML-based approaches.
 
\end{abstract}

% Note that keywords are not normally used for peerreview papers.

\section{Introduction}

High-Level Synthesis (HLS) fostered a revolution in hardware design. HLS frameworks allow the specification of hardware components in languages such as C, C++, or SystemC. As opposed to traditional Register Transfer Level (RTL) approaches, HLS flows do not require detailed descriptions of the logic gates, memory elements and interconnects comprising hardware implementations. Instead, these are automatically generated, based on the high-level specifications and on a set of directive values specifying optimizations such as the unrolling factor of loops and the inlining of functions. By decoupling specification from implementation, HLS allows unprecedented productivity, leading to considerable reductions in non-recurring engineering costs.

Nonetheless, while HLS allows to easily define vast design spaces for a given hardware specification, determining the performance (latency) and resource requirements (area, power) of each implementation still requires time-consuming syntheses. The amount of possible implementations of a design explodes exponentially with the number of applied directives, while, in general, only a few of them are Pareto-optimal from a performance/resources perspective. Exhaustive explorations are therefore wasteful (since only Pareto implementations are of interest) and impractical beyond very simple cases. 

Various strategies, which we summarize in Section \ref{sec:SoA}, have been proposed to identify (or approximate) the set of Pareto-implementations while minimising the number of synthesis runs \cite{LiuJun13} \cite{FerrettiJan18} \cite{ferretti2018lattice}. This problem is named \emph{HLS-driven Design Space Exploration (DSE)}. The proposed DSEs strategies are typically validated against exhaustive explorations, which the authors performed ad-hoc. Moreover, works such as \cite{wang20Mar} \cite{Ferretti20Oct} rely on prior knowledge to steer the HLS exploration process. 
Performing the huge number of synthesis required for validation or for generating a high-quality knowledge base entails a very high effort, which at present must be repeated ex-novo when investigating the performance of a novel DSE methodology. 

Against this backdrop, we introduce DB4HLS, a database of high-level synthesis design space explorations. The database comprises more than 100000 design points, reporting the synthesis outcomes of exhaustive explorations performed on 39 designs from the MachSuite \cite{reagen2014machsuite} benchmark suite. In addition, we define a simple domain-specific language to define design spaces, resulting in an open infrastructure that can be enriched by further contributions from the research community.

We believe that, by providing standardized synthesis data sets, our effort will allow easier comparisons among DSE strategies, enabling fairer evaluations of the strengths and weaknesses of each approach. It will also facilitate the development and assessment of future design exploration frameworks, spurring research in this challenging field.

\section{Related Works}

\label{sec:SoA}

%To tackle this problem researchers have proposed different strategies. 
State of the art DSE frameworks for HLS follow three main approaches. Black-box methodologies aim, after an initial phase, at iteratively refining explorations by smartly selecting additional design points. To this end, they employ unsupervised learning strategies such as clustering \cite{FerrettiJan18}, random forest \cite{LiuJun13}, lattice traversing \cite{ferretti2018lattice} and response surface models \cite{XydisOct14}.
Model-based strategies, on the other hand, estimate performance and resource requirements of implementations by developing an analytical formulation of the effect of directives when applied to a design. Typically, they can well approximate the Pareto set of best-performing implementations with few synthesis, but are restricted in the type of targeted optimizations (e.g., loop unrolling and dataflow in \cite{ZhongDec14}).
The authors of all these works adopt as figure of merit either the Hypervolume or the Average Distance from Reference Set (ADRS) for validation, and both require the computation of true Pareto frontiers from exhaustive explorations.
Recently, a promising research avenue has focused, instead, on exploiting prior knowledge in order to perform Design Space Exploration in hardware design. These works \cite{wang20Mar} \cite{Ferretti20Oct} leverage the availability of a comprehensive knowledge base, such as the one we describe in our paper, to achieve exploration results close to that of model-based strategies while being much more flexible in the number and type of supported directives.

While benchmark suites dedicated to hardware design are available, such as CHStone \cite{hara2008chstone}, MachSuite \cite{reagen2014machsuite}, Rosetta \cite{zhou2018rosetta} and S2CBench \cite{schafer2014s2cbench}, they only provide specifications (in the form C/C++ code) as benchmarks. Conversely, our DB4HLS suite offers rich and well-defined design spaces and related synthesis outcomes, greatly easing the burden of performing comparative evaluations of exploration methodologies.
To the best of our knowledge, this is the first database of HLS implementation made publicly available with the intent of standardize the evaluation process, and provide a source of knowledge for ML strategies.

\begin{table}[!tb]
% \footnotesize 
% \tiny
\centering
% \title{DNA source code encoding}
% \caption{\small List of functions explored from MachSuite\cite{reagen2014machsuite}, grouped by benchmark. The table reports: benchmark name, function name, and size of the configuration space ($|CS|$).}
\caption{\small{DSEs available in the database. Each entry reports benchmark, function name, and number of configurations ($|CS|$). All functions are from Machsuite\cite{reagen2014machsuite}}.}
\label{table:functions}
% \resizebox{\linewidth}{!}{%
\begin{tabular}{|l|l|l|}
% \hline
% \rowcolor{lightgray}
% \multicolumn{3}{|c|}{\textbf{DNA source code encoding}}                                                   \\
\hline
% \rowcolor{lightgray}
{\textbf{Benchmark}} & {\textbf{Function name}} & {\textbf{$|$\textbf{CS}$|$}} \\ 
\hline
spmv ellpack & ellpack & 1600\\
\hline
bfs bulk & bulk &2352 \\
\hline
md knn & md\_kernel & 1600\\
\hline
viterbi & viterbi &  1152\\
\hline
gemm ncubed & gemm &  2744\\
gemm blocked & bbgemm  & 1600\\
\hline
fft strided & fft  & 64\\
\hline
\multirow{2}{*}{sort merge} & ms\_mergesort & 4096\\
& merge & 4096\\
\hline
stencil stencil2d & stencil & 1344\\
stencil stencil3d & stencil3d & 1536\\
\hline
\multirow{7}{*}{radix sort} & update & 2400 \\
& hist & 4704\\
& init & 484 \\
& sum\_scan  & 1280\\
& last\_step\_scan  & 800\\
& local\_scan &  704\\
& ss\_sort  & 1792 \\
\hline
\multirow{7}{*}{aes} & aes\_addRoundKey  &  500\\
& aes\_subBytes  & 50\\
& aes\_addRoundKey\_cpy  & 625\\
& aes\_shiftRows & 20\\
& aes\_mixColumns & 18 \\
& aes\_expandEncKey & 216 \\
& aes256\_encrypt\_ecb & 1944\\
\hline
\multirow{13}{*}{backprop} & get\_delta\_matrix\_weights1 & 21952 \\
& get\_delta\_matrix\_weights2 & 31213 \\ 
& get\_delta\_matrix\_weights3 & 21952 \\
& get\_oracle\_activations1  & 2401 \\
& get\_oracle\_activations2  & 1372 \\
& product\_with\_bias\_input\_layer & 1372\\
& product\_with\_bias\_second\_layer & 686\\
& product\_with\_bias\_output\_layer & 392\\
& backprop & 2048 \\
& add\_bias\_to\_activations  & 1372\\
& soft\_max  & 64\\
& take\_difference & 512\\
& update\_weights & 1024\\
\hline
\end{tabular}
% }
\label{tab:all}
\vskip -3em
\end{table}

\section{Available design space explorations}
\label{sec:data}

We provide a rich set of DSEs by targeting the benchmarks of the MachSuite collection of designs~\cite{reagen2014machsuite}. We have performed DSEs for 39 out of 50 functions in the benchmark suite, discarding those having a variable latency due to input-dependent control flows, and those having very small design spaces. The considered functions present on average 40 lines of code, with the biggest having 308 lines of code. 

% For each design, we performed an extensive DSE across their configuration spaces -- the set of all possible implementation defined by the user -- up to tens of thousands of design points, these are reported in Table \ref{tab:all}.
% %The description of the configuration space considered for each function can be retrieved from the database from the corresponding entry in the \emph{Configuration Space} table.
% We used Vivado HLS~\cite{VivadoHLS} version 2018.2 to run syntheses with a target clock period of $10ns$ and targeting a ZynqMP Ultrascale+ (\texttt{\small xczu9eg}) FPGA chip.
% We constrained the design spaces sizes by employing only power-of-two values for directives having a numerical range (e.g., loop-unrolling and array-partitioning factors). 
% Moreover, in some DSEs we forced optimizations to have the same values, when intuitively such choice would lead to better cost/performance trade-offs (e.g., the partitioning factor of an array and the unrolling of a loop accessing it once every iteration).
We performed an exhaustive exploration of each design--according to the configuration space defined by the user--running more than 100000 synthesis. Table \ref{tab:all} lists all the designs explored and their configuration space size.

We used Vivado HLS~\cite{VivadoHLS} version 2018.2 to perform the syntheses
, and we targeted
%LP  . Moreover, we synthesised the designs targeting 
a ZynqMP Ultrascale+ (\texttt{\small xczu9eg}) FPGA chip, with a target clock of $10ns$.

To restrain the design spaces sizes, we have constrained directive set values with a numerical range  (e.g., the unrolling factor) to power-of-two or integer divisor of the maximum admissible values (e.g., number of loop iterations).
Moreover, for some designs, different optimizations are forced to have the same values when intuitively such choice would lead to better cost/performance trade-offs (e.g., binding the loop unrolling factor to the array partitioning one).

% Even when considering these constraints, the data collection required more than 4 years of single-core machine time. To speed up this process, GNU Parallel was adopted to collect synthesis data from up to 60 parallel instances of Vivado HLS, allowing us to populate the database in approximately 25 days of wall-clock time.
Even when considering these constraints, the data collection required more than 4 years of single-core machine time. To speed up this process, GNU Parallel was adopted to collect synthesis results from 60 parallel Vivado HLS instances, allowing us to populate the database in approximately 25 days of wall-clock time.

\begin{figure}[t]
\centering
\includegraphics[width=\linewidth]{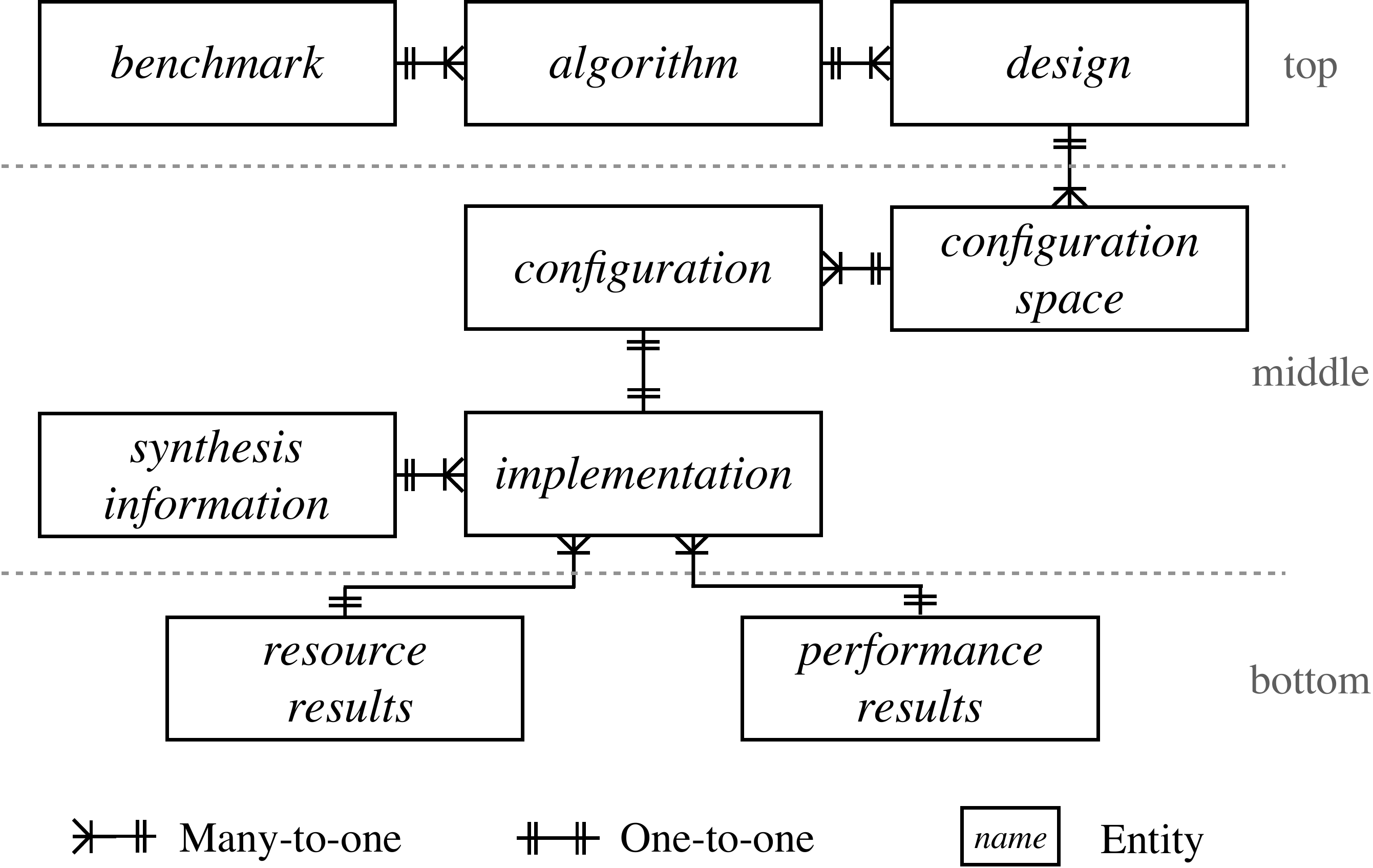}
\caption{\small Simplified scheme of the Entity-Relationship Diagram (ERD) of the DB4HLS syntheses database.}
\label{fig::erd}
\vskip -1em
\end{figure}

\section{DB4HLS infrastructure}
\label{sec:implementation}

In addition to the DSE data, the DB4HLS framework offers a) a database infrastructure hosting DSE in a structured and easy-to-access way, b) a domain-specific language used to describe a configuration space for a target design, c) an interface to generate new explorations and further enrich the database. The remaining of this section describes these further contributions in details.

\subsection{A database for DSEs}
\label{subsec:DB}

% The implemented Database, implemented in MySQL, comprises a description of the application targeted for exploration (upper part of Figure \ref{fig::erd}), and that of the explored HLS optimizations applied to each application (middle part of Figure \ref{fig::erd}). Finally, it reports the  resource and performance results obtained by synthesis (as described in the bottom part of the figure). Each of these components is described more in detail in the following paragraphs.

\begin{figure*}[t]
%\centering

\begin{minipage}{.48\textwidth}
  \centering
%  \vskip -2em
\begin{lstlisting}[caption={\texttt{last\_step\_scan} design (C code).}, label={code:last_step_scan} ,linewidth=\columnwidth,breaklines=true,language=C,
    firstnumber=1,
    numbers=left,
    stepnumber=1,
    numberblanklines=false,
    tabsize=1,
    basicstyle=\small,
    % basicstyle=\fontsize{8}{10}\ttfamily,
    breakatwhitespace=false,
    xleftmargin=2em]
void last_step_scan(int bucket[SIZE], int sum[RADIX]){
  int i, j, k;
  loop_1:for(i = 0; i < RADIX;i++){
    loop_2:for(j = 0; j < BLOCK; j++) {
      k = (i * BLOCK) + j;
      bucket[k] = bucket[k] + sum[i];
    }
  }
}
\end{lstlisting}
\end{minipage}
\hfill
\begin{minipage}{.48\textwidth}
%\vskip 0.5em
  \centering
\begin{lstlisting}[caption=Configuration Space of \texttt{last\_step\_scan}.,label={sweep:last_step_scan},linewidth=\columnwidth,breaklines=true,language=C,
    firstnumber=1,
    numbers=left,
    stepnumber=1,
    numberblanklines=false,
    tabsize=1,
    basicstyle=\small,
    % basicstyle=\fontsize{8}{10}\ttfamily,
    breakatwhitespace=false,
    xleftmargin=2em]
resource;last_step_scan;bucket;{RAM_2P_BRAM}
resource;last_step_scan;sum;{RAM_2P_BRAM}
array_partition;last_step_scan;bucket;1;{cyclic,block};{1->512,pow_2}
array_partition;last_step_scan;sum;1;{cyclic,block};{1->128,pow_2}@bind_a
unroll;last_step_scan;last_1;{1->128,pow_2}@bind_a
unroll;last_step_scan;last_2;{1,2,4,8,16}
clock;{10}
\end{lstlisting}
% \vskip -2em
\end{minipage}
% \vskip -2em
% \end{mdframed}
\vskip -1em
\caption{\small Left: Snippet of the \texttt{\small last\_step\_scan} C code function from MachSuite\cite{reagen2014machsuite}. 
We rewrote the code to increase the readability without affecting its functionality.
% The code was  rewritten  to  increase  readability,  without  modifying  its functionality.
Right: An associated Configuration Space Descriptor (CSD). }\label{fig:mot_example}
\vskip -1.5em
\end{figure*}

The database structure, implemented in MySQL, comprises a description of the design targeted for exploration (top part of Figure \ref{fig::erd}), and that of the explored HLS optimizations applied to each design (middle part of Figure \ref{fig::erd}). Finally, it reports the resource and performance results obtained by synthesis (as described in the bottom part of the figure). Each of these components is described more in detail in the following.

Similarly to the taxonomy adopted in MachSuite \cite{reagen2014machsuite}, applications are identified by the \emph{benchmark} they belong to (e.g.: \texttt{aes}), by the \emph{algorithm} they realize (e.g.: \texttt{aes256\_encrypt}) and by the \emph{design} implementing such algorithms. As an example, two variants are provided by MachSuite for the \texttt{aes256\_encrypt} algorithm (one using lookup tables to store  encryption keys and one generating the values online), each corresponding to a separate design specified as C++ code.

A descriptor of the HLS optimizations considered for the DSEs are stored as entries in \emph{configuration space} table. Multiple explorations (hence, rows in the configuration space table) for the same design are possible, corresponding to different choices of optimizations, or explorations targeting different tools/FPGAs, or even contributions from different researchers. An entry in the \emph{configuration space} table is linked to many entries of the \emph{configuration} table, where each entry indicates a specific element of the design space.

A line in the \emph{configuration} table (that indicates the set of HLS optimizations defining a design space element) is linked to an entry in the \emph{implementation} table. Furthermore, the \emph{synthesis information} table provides additional information on each performed synthesis: the synthesis timestamp, the contributor that originated the data, the employed synthesis tool and version, and the targeted FPGA. Finally, each implementation links to one or more entries in the \emph{resources} and \emph{performance} tables, which report the synthesis outcomes. Resources are expressed as employed Flip-Flops, Look-Up Tables, Block RAMs (BRAM) and DSP blocks, while performances are reported in terms of effective latency.

\subsection{A domain-specific language for DSEs}\label{subsec:dsl}

Generating the different configurations associated with an DSE is a tedious and error-prone process when performed by hand. We therefore developed a Domain-Specific Language (DSL) to automatically and concisely define configuration spaces by employing Configuration Space Descriptors (CSDs). 

Each line of a descriptor encodes a \emph{knob}, which comprises a directive type, a label corresponding to its location in the design C/C++ code, and one or multiple sets of values. The number of sets is equal to the number of parameters required by the directive type. Values can be numerical when expressing optimizations such as loop unrolling or array partitioning factors, or categorical when determining the type of employed FPGA resources such as BRAM types.
A shorthand is provided for expressing regular value series (e.g., a succession of power-of-two values). Finally, we provide a \texttt{\small{@bind}} decorator, which constraints the values associated with different directives.
% forces two or more directives of a configuration to use the same associated directive value.

% Figure \ref{fig:mot_example} shows for the function \texttt{\small last\_step\_scan} (Snippet \ref{code:last_step_scan}) an example of DSL defined to describe the configuration space defined for its DSE (Snippet~\ref{sweep:last_step_scan}).
% The DSL defines seven different knobs. Line 1 of Snippet~\ref{sweep:last_step_scan} shows a knob with a single value: it associates a dual-ported BRAM to the array \texttt{\small bucket} that is the input of the function. Similarly, line 2 defines a dual-ported BRAM for the array \texttt{\small sum}.
% Line 3 instead defines a knob governing the array\_partitioning directive defined by all the pairs having one of two partitioning strategies (\texttt{\small cyclic} and \texttt{\small block}) as the first component, and the ten possible partitioning factors  (all the powers of two from 1 up to 512) as the second one. The same is done in line 4, but defining a different set of partitioning factors (all the powers of two from 1 up to 128). 
Figure \ref{fig:mot_example} shows, for the \texttt{\small last\_step\_scan} function in Snippet \ref{code:last_step_scan}, an example of DSL descriptor created to define its configuration space (Snippet~\ref{sweep:last_step_scan}) created using the DSL.
The DSL descriptor defines seven different knobs. Lines 1 and 2 of Snippet~\ref{sweep:last_step_scan} show two knobs associating a dual-port BRAM to the input array \texttt{\small bucket}, and \texttt{\small sum} respectively.
Lines 3 and 4 define knobs specifying the array\_partitioning directive. These directives are created as combinations of partitioning strategies and partitioning factors.
Both line 3 and 4 combine two partitioning strategies (\texttt{\small cyclic} and \texttt{\small block}) with the associated directive values set for the partitioning factors--all the powers of two from 1 up to 512 for knob 3, and all the powers of two from 1 up to 128 for knob 4.
Then line 5 and 6 define for \texttt{loop\_1} and \texttt{loop\_2} the associated set of unrolling factors to consider during the exploration, all the powers of two from 1 up to 128 and 16, respectively.
Both line 4 and 5 have a binding decorator (\texttt{\small @bind\_a}), that specifies that the array partitioning directive and the unrolling one must have the same partitioning and unrolling factor for all the configurations described by the CSD. 
Finally line 7 defines the target clock.

%\demo

% The DSL generates the set of configurations of the design space as the Cartesian product of all knob values: $CS = K_{1} \times K_{2} \times \dots \times K_{N}$;
% % \begin{equation}
% % CS = K_{1} \times K_{2} \times \dots \times K_{N}
% % \end{equation}
% where  $N$ is the number of considered knobs, and $K_{i}$ is the set of values related to each $i$ knob, i.e. the set of values that the directive associated to the knob $i$ can assume,
% taking into account the restrictions imposed by the \texttt{bind} decorator. For a directive  with  multiple parameters, $K_{i}$ is itself the Cartesian product among each set of values.
% The size of the configuration space is then given by its cardinality ($|CS|$).

The configuration space resulting from a DSL descriptor having $N$ different knobs is the Cartesian product of all knob values: $CS = K_{1} \times K_{2} \times \dots \times K_{N}$;
where $K_{i}$ is the directive values set related to knob $i$,
taking into account the restrictions imposed by the \texttt{bind} decorator. In case of directives requiring multiple parameters, the knob $K_{i}$ is itself the Cartesian product among each set of values associated to the knob.
Lastly, the total number of configurations, i.e., the configuration space size, is given by its cardinality ($|CS|$).

The configuration space descriptor of Figure \ref{fig:mot_example} in Snippet \ref{sweep:last_step_scan} describes a configuration space with 1600 different configurations. Without the binding decorator, the cardinality of the configuration space would be 12800.

\subsection{A framework for parallelizing HLS runs}

\begin{figure}[t]
\centering
\includegraphics[width=\linewidth]{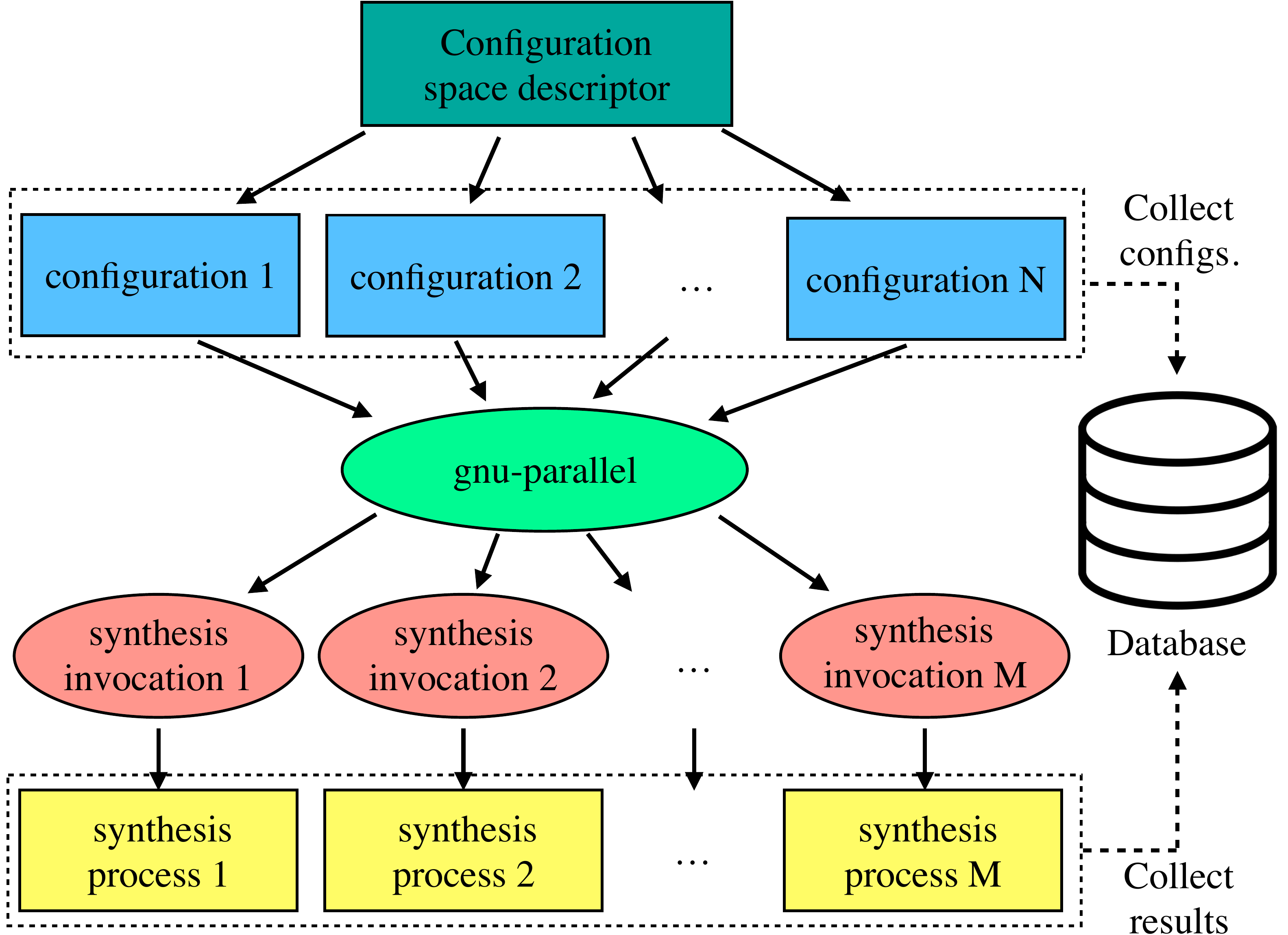}
\caption{\small Adding DSE to DB4HLS with parallel processing.}
\label{fig::infrastructure}
\vskip -1.5em
\end{figure}

Figure \ref{fig::infrastructure} gives a high-level view of the infrastructure, realized through Bash and Python scripts, which we provide to automate DSE and commit their outcomes in DB4HLS. Starting from a user-provided design and Configuration Space Descriptor (CSD), configuration files are automatically generated and stored in the database. Then, using GNU Parallel \cite{gnuparallel}, a tunable number of instances of an employed HLS tool (we use Vivado HLS for the data collection described in Section \ref{sec:data}) are concurrently and independently executed, one for each configuration. As synthesis runs terminate, the retrieved performance and resources information are also stored in DB4HLS, and new HLS processes are launched until all configurations have been explored. 

MySQL statements can then be used to retrieve data from the tables in the database and to access the design's implementations and the associated performance and resources results.

\section{Case Study}

\begin{figure}[t]
\centering
% \begin{minipage}{.5\linewidth}
  \centering
%   \captionsetup{justification=centering,margin=2mm}
  \captionsetup{margin=1mm}
  \includegraphics[width=\linewidth]{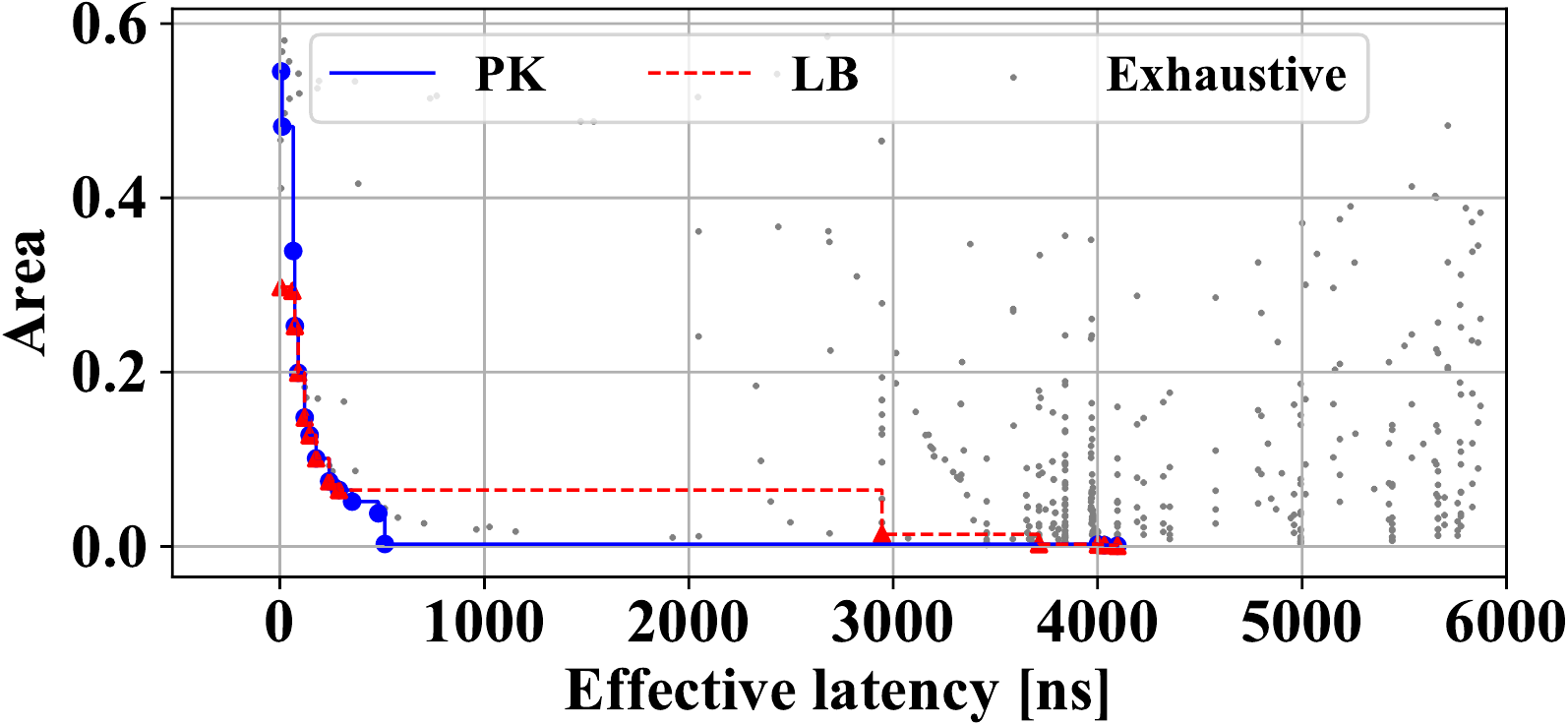}
% \end{minipage}%
% % \begin{minipage}{.5\linewidth}
%   \centering
% \captionsetup{margin=1mm}
%   \includegraphics[width=\linewidth]{imgs/bad-AEsim.pdf}
% \end{minipage}
%  \caption{\small Grey dots represent the area and latency of the 704 implementations belonging to the \texttt{\small local\_scan} design space provided by DB4HLS. The figure also reports the approximate Pareto fronts retrieved by the lattice methodology described in \cite{ferretti2018lattice} (LB) and by the prior-knowledge strategy in \cite{Ferretti20Oct} (PK).}
 \caption{\small Example of DSEs comparison employing DB4HLS.}\label{fig:exploration}
%  both to allow a direct comparison among two different methodologies (LB \cite{ferretti2018lattice} and PK \cite{Ferretti20Oct}) and as initial source of knowledge for PK. }\label{fig:exploration}
% \vspace{-5mm}
\vskip -1.5em
\end{figure}

Herein, we showcase two possible uses of DB4HLS. We use the database both to compare the results of two DSE methodologies, and as a source of knowledge for one of them. We employed a lattice-based strategy (LB) from \cite{ferretti2018lattice}, and one leveraging prior knowledge (PK)\cite{Ferretti20Oct}, to perform DSEs for the \texttt{\small local\_scan} design space available in DB4HLS.
Figure \ref{fig:exploration} reports the Pareto curve obtained by LB and PK for the \texttt{\small local\_scan} design space. Grey dots represent the area and latency of the 704 implementations belonging to the \texttt{\small local\_scan} design space provided by DB4HLS. The figure also reports the approximated Pareto fronts retrieved by the lattice methodology described in \cite{ferretti2018lattice} (LB) and by the prior-knowledge strategy in \cite{Ferretti20Oct} (PK).
% In this example the database has a double role. It is used both to allow a direct comparison among two different methodologies and as initial source of knowledge for the PK one. 

In this scenario, DB4HLS is employed to comparatively evaluate the two strategies, without requiring to re-run ex-novo a large number of time-consuming synthesis runs. Besides, for PK, the database mandates the availability of a set of \emph{source} design spaces in order to extract previous knowledge. In fact, DB4HLS can be effectively employed in these cases, or in similar ML-based methods \cite{wang20Mar}, to provide the required knowledge base. 

\section{Conclusions}

DB4HLS offers an extensive set of DSEs targeting functions from MachSuite \cite{reagen2014machsuite}. The data collection is made publicly available and will be will be updated increasing the number of design explorations and targeted benchmarks. In addition, further design spaces can be effectively defined through a novel domain-specific language and a framework to efficiently contribute novel explorations to DB4HLS. Both the DB4HLS database and the framework for DSE generation are publicly available at \small{\texttt{https://www.db4hls.inf.usi.ch/}}. 

\ifCLASSOPTIONcaptionsoff
  \newpage
\fi

% trigger a \newpage just before the given reference
% number - used to balance the columns on the last page
% adjust value as needed - may need to be readjusted if
% the document is modified later
%\IEEEtriggeratref{8}
% The "triggered" command can be changed if desired:
%\IEEEtriggercmd{\enlargethispage{-5in}}

% references section

% can use a bibliography generated by BibTeX as a .bbl file
% BibTeX documentation can be easily obtained at:
% http://mirror.ctan.org/biblio/bibtex/contrib/doc/
% The IEEEtran BibTeX style support page is at:
% http://www.michaelshell.org/tex/ieeetran/bibtex/
%\bibliographystyle{IEEEtran}
% argument is your BibTeX string definitions and bibliography database(s)
%\bibliography{IEEEabrv,../bib/paper}
%
% <OR> manually copy in the resultant .bbl file
% set second argument of \begin to the number of references
% (used to reserve space for the reference number labels box)
%\begin{thebibliography}{1}

%\bibitem{IEEEhowto:kopka}
%H.~Kopka and P.~W. Daly, \emph{A Guide to \LaTeX}, 3rd~ed.\hskip 1em plus
%  0.5em minus 0.4em\relax Harlow, England: Addison-Wesley, 1999.
\bibliographystyle{IEEEtran}
\bibliography{biblio} 

%\end{thebibliography}

% biography section
% 
% If you have an EPS/PDF photo (graphicx package needed) extra braces are
% needed around the contents of the optional argument to biography to prevent
% the LaTeX parser from getting confused when it sees the complicated
% \includegraphics command within an optional argument. (You could create
% your own custom macro containing the \includegraphics command to make things
% simpler here.)
%\begin{IEEEbiography}[{\includegraphics[width=1in,height=1.25in,clip,keepaspectratio]{mshell}}]{Michael Shell}
% or if you just want to reserve a space for a photo:

%\begin{IEEEbiography}{Michael Shell}
%Biography text here.
%\end{IEEEbiography}
%
%% if you will not have a photo at all:
%\begin{IEEEbiographynophoto}{John Doe}
%Biography text here.
%\end{IEEEbiographynophoto}
%
%% insert where needed to balance the two columns on the last page with
%% biographies
%%\newpage
%
%\begin{IEEEbiographynophoto}{Jane Doe}
%Biography text here.
%\end{IEEEbiographynophoto}

% You can push biographies down or up by placing
% a \vfill before or after them. The appropriate
% use of \vfill depends on what kind of text is
% on the last page and whether or not the columns
% are being equalized.

%\vfill

% Can be used to pull up biographies so that the bottom of the last one
% is flush with the other column.
%\enlargethispage{-5in}

% that's all folks
\end{document}